\documentclass{iau} 
\usepackage{graphicx}

\usepackage{url}

\title[Searching for dual AGN in J1425+3231] 
{Searching for a pair of accreting supermassive black holes in J1425+3231}

\author[K. \'E. Gab\'anyi et al.]
{K. \'E. Gab\'anyi$^{1,2}$,
S. Frey$^{1,2}$, Z. Paragi$^3$, T. An$^4$
 \and S. Komossa$^5$}

\affiliation{$^1$F\"OMI, Satellite Geodetic Observatory, P.O. Box 585, 1592 Budapest, Hungary \\ email: {\tt krisztina.g@gmail.com} \\[\affilskip]
$^2$Konkoly Observatory, MTA CsFK, P.O. Box 67, 1525 Budapest, Hungary\\ 
$^3$Joint Institute for VLBI ERIC, Postbus 2, 7990 AA Dwingeloo, the Netherlands \\
$^4$Shanghai Astronomical Observatory, CAS, 80 Nandan Road, 200030 Shanghai, P.R. China\\
$^5$QianNan Normal University for Nationalities, Longshan Street, Duyun City of Guizhou Province, P.R. China}

\pubyear{2016}
\volume{324}  
\jname{New Frontiers in Black Hole Astrophysics}
\editors{A.C. Editor, B.D. Editor \& C.E. Editor, eds.}
\begin{document}

\maketitle

\begin{abstract}
In hierarchical structure formation scenarios, merging galaxies are expected to be seen in different phases of their coalescence. Simulations suggest that simultaneous activity of the supermassive black holes (SMBHs) in the centres of the merging galaxies may be expected at kpc-scale separations. Currently, there are no direct observational methods which allow the selection of a large number of such dual active galactic nuclei (AGN) candidates. SDSS J142507.32+323137.4 was reported as a promising candidate source based on its optical spectrum. Here we report on our sensitive e-MERLIN observations performed at $1.6$ and at $5$\,GHz, which show that the optical spectrum of the source can be more straightforwardly explained with jet--cloud interactions instead of the dual AGN scenario.
\keywords{galaxies: active, quasars: individual: SDSS J142507.32+323137.4, radio continuum: galaxies, techniques: interferometric}
\end{abstract}
\firstsection 
\section{Introduction}

Hierarchical structure formation models predict that galaxies and their central supermassive black holes (SMBHs) form and grow via merging events. These events may increase activity of the central SMBHs and/or star formation in the host galaxies. Simulations suggest that simultaneous activity of the black holes may be expected at kpc-scale separations (\cite[Van Wassenhove et al., 2012]{vanWassenhove}). Additionally, according to \cite[Steinborn et al. (2015)]{Steinborn} and \cite{volonteri}, the number of luminous dual AGN increases with decreasing separation of the supermassive black holes. Directly resolving the tightest AGN pairs is only feasible in the most nearby Universe with the current optical and X-ray instrumentation, while cosmological simulations of \cite{volonteri} show that less dual AGN are expected at low redshift.

At radio wavelengths, the technique of very long baseline interferometry (VLBI) provides the highest achievable angular resolution, however only $\approx 10$\,\% of AGN are radio-loud. Since a large fraction of low-luminosity AGN are very compact on milliarcsecond resolution (\cite[Panessa \& Giroletti, 2013]{panessa}), 
VLBI is a powerful tool to confirm such AGN candidates. But low-luminosity AGN may have very resolved structures on parsec scales as well (e.g., \cite[Nagar et al., 2002]{nagar}), which can only be imaged by arrays with shorter baselines.
Also VLBI instruments are not ideal for large surveys due to their limited field of view. Therefore, it is essential to assemble reliable samples of candidate dual AGN. Currently, however, no method is known which can be used to select such sources. Searching for double-peaked narrow optical emission lines emerging from the two separate narrow-line regions (NLR) of dual AGN was proposed as such a method (e.g. \cite[Wang et al., 2009]{double}). However, the double-peaked spectral lines can also arise due to other effects occurring in a single AGN, e.g. peculiar kinematics or jet--cloud interactions (\cite[Heckman et al., 1984]{heckman}), a rotating, disk-like NLR (\cite[Xu \& Komossa, 2009]{Xu}), and the combination of a blobby NLR and extinction effects (\cite[Crenshaw et al., 2010]{Crenshaw}). Therefore it is important to follow-up those radio-emitting sources claimed to be promising dual AGN candidates using VLBI technique. The SMBHs in dual AGN, once they finally merge and coalesce, are the loudest sources of gravitational wave emission in the universe. Observing active SMBHs in all stages of merging from wide to small separation is therefore of great interest.

SDSS\,J142507.32+323137.4 (hereafter J1425+3231, $z=0.478$) was reported as a promising dual AGN candidate by \cite[Peng et al. (2011)]{Peng} based upon its double-peaked narrow optical emission lines. \cite[Frey et al. (2012)]{Frey} found two compact radio emitting features at a projected separation of $\sim 2.6$\,kpc at $1.7$\,GHz in a European VLBI Network (EVN) observation. At $5$\,GHz, only the brighter feature was detected with EVN. Below we report on our e-Multiple Element Remotely Linked Interferometer Network (e-MERLIN) dual-frequency observations of J1425+3231 (project: CY2205), which were aimed at finding out whether the two detected radio features are the cores of two radio-emitting AGN, or they belong to one AGN, and represent core and extended features.

\section{Observations and Data Reduction}

The 1.6 GHz e-MERLIN observations took place between 2015 October 2 and 4, the interferometric array consisted of the following antennas: Jodrell Bank Mk2, Defford, Pickmere, Darnhall, and Cambridge. The on-source integration time was $27$ h. The 5 GHz e-MERLIN observations took place on 2014 November 1 and 11, the array consisted of: Jodrell Bank Mk2, Cambridge, Defford, Pickmere, and Knockin. The on-source integration time was $19$ h. In both cases the phase reference source was J1422+3223; this source was used as phase reference source also in the EVN observations. 

Data reduction was done using the National Radio Astronomy Observatory (NRAO) Astronomical Image Processing System (AIPS, \cite[Greisen, 2003]{aips}), following the e-MERLIN cookbook\footnote{\url{http://www.e-merlin.ac.uk/data_red/tools/e-merlin-cookbook_V3.0_Feb2015.pdf}}, and with help from the e-MERLIN science team (R. Beswick).

We used the modelfit task of the Difmap package \cite[(Shepherd et al., 1994)]{difmap} to model the brightness distribution with circular Gaussian components. The full-width at half maximum (FWHM) sizes, flux densities, and projected separations from component C of the fitted Gaussian components are given in Table \ref{tab:flux}. 

\begin{table}
	\begin{center}
		\caption{Details of the fitted circular Gaussian components to the e-MERLIN observations of J1425+3231.} 
		\label{tab:flux}
		{\scriptsize
  		\begin{tabular}{|c|c|c|c|c|}\hline 
		{\bf $\nu$} & {\bf ID} & {\bf Position} & {\bf Flux density} & {\bf FWHM}\\
{\bf (GHz)} & & {\bf (mas)} & {\bf ($\mu$Jy)} & {\bf (mas)}\\
		\hline
		 1.6 & C & -- & $712 \pm 63$ & $33 \pm 2$ \\
		  & S & 428 & $1122 \pm 81$ & $104 \pm 6$ \\
		  & N & 202 & $1835 \pm 116$ & $204 \pm 12$ \\
		  \hline
		  5 & C & -- & $319 \pm 57$ & $38 \pm 6$ \\
		  \hline		   
		\end{tabular}
		}
	\end{center}
\end{table}

\section{Results}

\begin{figure}
	\begin{minipage}{\columnwidth}
		\centering
		\includegraphics[width=0.49\columnwidth]{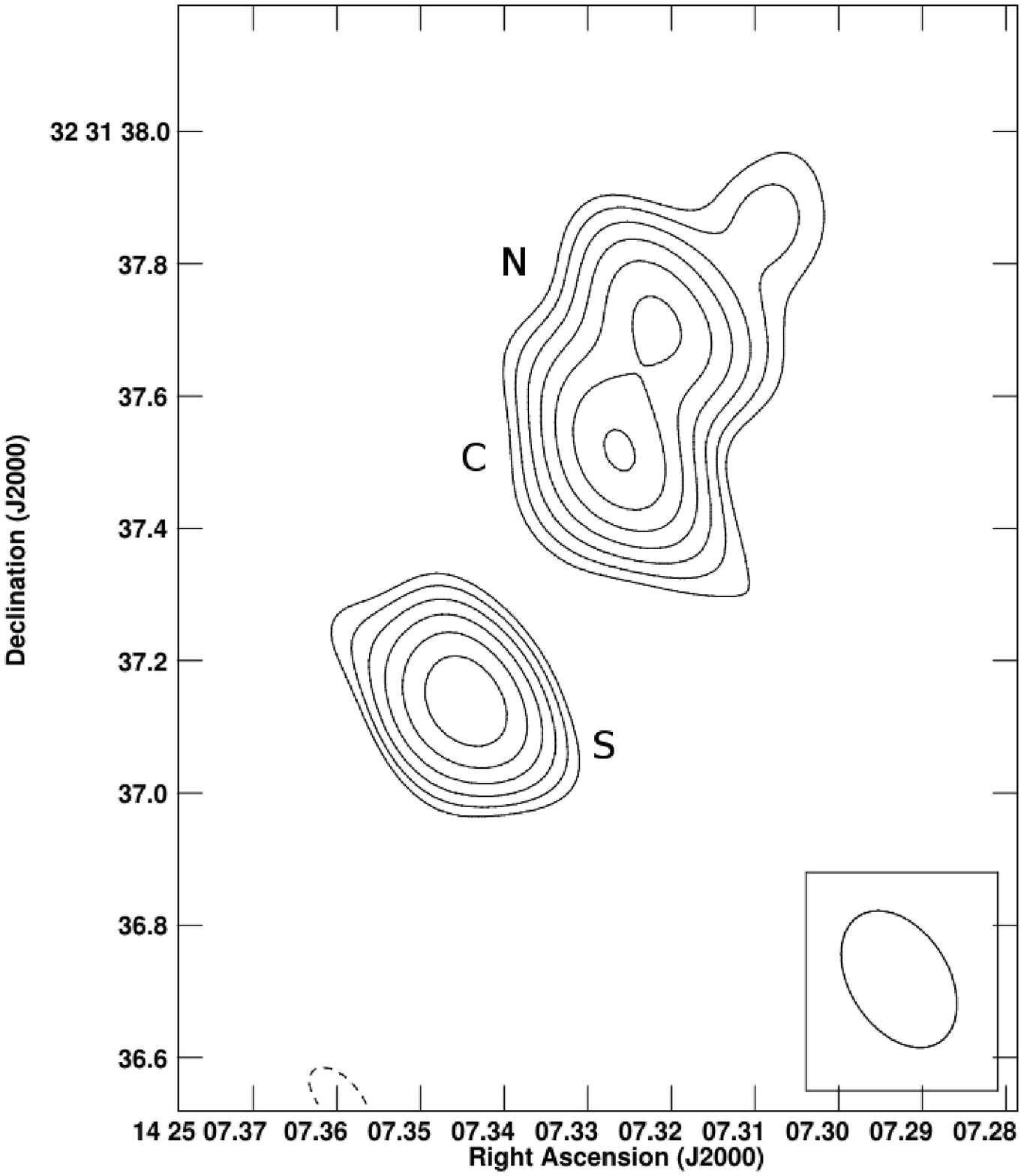}
		\includegraphics[width=0.49\columnwidth]{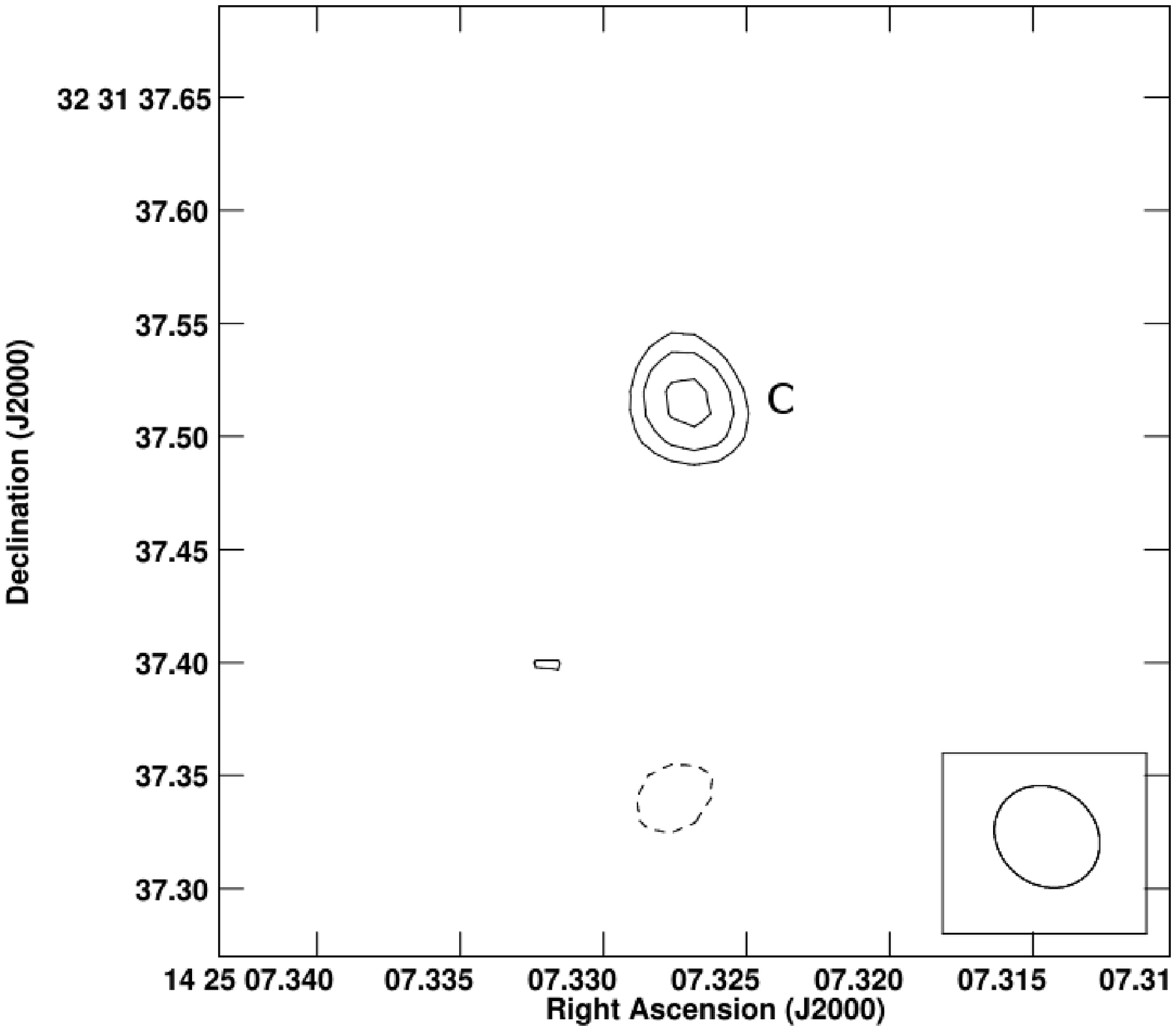}
	\end{minipage}
	\caption{{\it Left panel:} 1.6-GHz e-MERLIN image of J1425+3231. The peak brightness is $837\mu$Jy\,beam$^{-1}$. The lowest positive contour level is at $100\mu$Jy\,beam$^{-1}$ ($4\sigma$ image noise level), further contour levels increase by a factor of $\sqrt{2}$. The beamsize is $226 \times 149$ mas at a position angle of $33^\circ$ and shown in the lower right corner of the image. {\it Right panel:} 5-GHz e-MERLIN image of J1425+3231. The peak brightness is $154\mu$Jy\,beam$^{-1}$. The lowest positive contour level is at $69\mu$Jy\,beam$^{-1}$ ($4\sigma$ image noise level), further contour levels increase by a factor of $\sqrt{2}$. The beamsize is  $49 \times 43$ mas at a position angle of $52^\circ$ and shown in the lower right corner of the image.}
	\label{fig:eMERLIN}
\end{figure}

At 1.6 GHz, J1425+3231 shows a rich structure in the e-MERLIN observation (see left panel of Fig. \ref{fig:eMERLIN}). We detected component C and S also seen in the 1.7 GHz EVN image, and additionally a bright northern feature, component N. At 5 GHz, similar to the EVN results, we detected only component C (see right panel of Fig. \ref{fig:eMERLIN}). There is no additional radio emission in the 5-GHz image down to $\sim 100\mu$Jy/beam ($6\sigma$ noise level).

The sum of the flux densities of the components detected at $1.6$\,GHz ($3.67 \pm 0.16$ mJy) is in good agreement with the flux density measured for the source at the close frequency of $1.4$\,GHz in the lower resolution observation conducted with the Very Large Array ($3.28 \pm 0.14$\,mJy) within the framework of the Faint Images of the Radio Sky at Twenty-centimeters (FIRST, \cite[Becker et al., 1995]{first}) survey. The e-MERLIN observations thus recovered all arcsec-scale radio emission in the source measured in the FIRST survey. The slightly larger e-MERLIN flux density can be caused by source variability. 

Compared with the 1.7-GHz EVN data, both detected components (C and S) show larger flux densities in the e-MERLIN observation. This can indicate that at this frequency, e-MERLIN picked up the large-scale structure resolved out in the EVN observation and/or it may be caused by flux density variability. However, the large FWHM size of component S ($104 \pm 6$\,mas) indicates that the resolution effect is more likely in that case. At $5$\,GHz, the flux densities of the only detected component (component C) agree within the uncertainties in the e-MERLIN and EVN observations. 

Components N and S have significantly larger FWHM sizes ($104 \pm 6$\,mas, and $204 \pm 12$\,mas, respectively) in the e-MERLIN data than component C ($33 \pm 2$\,mas). This suggests that component C is a compact core, while components N and S are more extended features probably associated with the jet or lobe of the AGN. On the other hand, the detection of emission at the position of component S with the EVN shows that it contains a compact radio-emitting feature.

There are two scenarios that can explain the radio structure of J1425+3231. (i) There is a single low-power radio-emitting nucleus in the source at the position of component C, the synchrotron-emitting jet base. A jet-like feature (component N) to the north, and a lobe (component S) to the south could be imaged with e-MERLIN at $1.6$\,GHz. The latter contains a compact hotspot, which was detected with the EVN. However, component N was resolved out with EVN completely. In this scenario, the double-peaked optical emission lines are related to the jet-driven outflows and not to dual AGN. A similar observational character has been found in other radio AGN with double-peaked narrow optical lines (e.g. 3C\,316 An et al., 2013). (ii) There are two low-power radio emitting nuclei in the source at the position of components C and S. Their compact cores were detected in the EVN observation at $1.7$\,GHz. One of them (C) has a jet-like extension to the north (N). The large-scale structure of the secondary nucleus (S) was detected with e-MERLIN. The second option is less likely since S was not detected at any of the higher-frequency radio observations, indicating a very steep radio spectrum ($\alpha<-2$, the spectral index is defined as $S\sim \nu^\alpha$), which is at odds with its identification as an AGN core.

\section{Summary}

J1425+3231 was proposed to be a dual AGN candidate because of its double-peaked optical spectrum (\cite[Peng et al., 2011]{Peng}). High-resolution radio observations with the EVN showed that it has two compact radio emitting feature. Our follow-up e-MERLIN observations revealed a rich morphology in the source at $1.6$\,GHz, but detected only one radio-emitting feature at $5$\,GHz. While the scenario involving dual radio-emitting AGN cannot be completely ruled out, our results are more indicative of one low-luminosity AGN with two-sided jet and/or lobe-like features in J1425+3231.

\smallskip
The e-MERLIN is a National Facility operated by the University of Manchester at Jodrell Bank Observatory on behalf of the UK Science and Technology Facilities Council (STFC). This research was supported by the Hungarian National Research, Development and Innovation Office (OTKA NN110333) and by the China--Hungary Collaboration and Exchange Programme by the International Cooperation Bureau of the Chinese Academy of Sciences.






\begin{thebibliography}{}

\bibitem[An et al. (2013)]{3C316}
{An, T., Paragi, Z., Frey, S., et al.} 2013
\textit{MNRAS}, 433, 1161

\bibitem[Becker et al. (1995)]{first}
{Becker, R. H., White, R. L., Helfand, D. J.} 1995,
\textit{ApJ}, 450, 559

\bibitem[Crenshaw et al. (2010)]{Crenshaw}
{Crenshaw, D. M., Schmitt, H. R., Kraemer, S. B., et al.} 2010,
\textit{ApJ}, 708, 419

\bibitem[Frey et al. (2012)]{Frey}
{Frey, S., Paragi, Z., An, T., Gab\'anyi, K. \'E.} 2012,
\textit{MNRAS}, 425, 1185

\bibitem[Greisen (2003)]{aips}
{Greisen, E. W.} 2003,
\textit{Information Handling in Astronomy - Historical Vistas}, 285, 109

\bibitem[Heckman et al. (1984)]{heckman}
{Heckman, T. M., Miley, G. K., Green R. F.} 1984,
\textit{ApJ}, 281, 525

\bibitem[Helfand et al. (2015)]{firstdata}
{Helfand, D. J., White, R. L., Becker, R. H.} 2015,
\textit{ApJ}, 801, 26

\bibitem[Nagar et al. (2002)]{nagar}
{Nagar, N. M., Falcke, H., Wilson, A. S., et al.} 2002,
\textit{A\&A}, 392, 53

\bibitem[Panessa \&Giroletti (2013)]{panessa}
{Panessa, F., Giroletti, M.} 2013,
\textit{MNRAS}, 432, 1138

\bibitem[Peng et al. (2011)]{Peng}
{Peng, Z.-X., Chen, Y., Gu, Q., et al.} 2011,
\textit{Research in Astronomy and Astrophysics}, 11, 411

\bibitem[Shepherd et al. (1994)]{difmap}
{Shepherd, M. C., Pearson, T. J., Taylor, G. B.} 1994,
\textit{BAAS}, 26, 987

\bibitem[Steinborn et al. (2015)]{steinborn}
{Steinborn, L. K., Dolag, K., Hirschmann, M., et al.} 2015,
\textit{MNRAS}, 458, 1013

\bibitem[Van Wassenhove et al. (2012)]{vanWassenhove}
{Van Wassenhove, S., Volonteri, M., Mayer, L., et al.} 2012,
\textit{ApJ} (Letters) 748, L7

\bibitem[Volonteri et al. (2016)]{volonteri}
{Volonteri, M., Dubois, Y., Pichon, C., Devriendt, J.} 2016,
\textit{MNRAS}, 460, 2979

\bibitem[Wang et al. (2009)]{double}
{Wang, J.-M., Chen, Y.-M., Hu, C., et al.} 2009,
\textit{ApJ} (Letters) 705, L76

\bibitem[Xu \& Komossa (2009)]{Xu}
{Xu, D., Komossa, S.} 2009,
\textit{ApJ} (Letters) 705, L20


\end{thebibliography}
\end{document}